\documentclass[conference]{IEEEtran}
\IEEEoverridecommandlockouts
\usepackage{amsmath,epsfig}
\usepackage{bm}
\usepackage{latexsym}
\usepackage{amsmath}
\usepackage{amssymb}
\usepackage{graphicx}

\def\bE{\mathbb{E}}

\def\bI{\mathbf{I}}

\def\bR{\mathbf{R}}

\def\bT{\mathbf{T}}
\def\bw{\mathbf{w}}

\def\bv{\mathbf{v}}
\def\be{\mathbf{e}}
\def\bq{\mathbf{q}}
\def\bh{\mathbf{h}}
\def\bg{\mathbf{g}}
\def\bG{\mathbf{G}}

\def\ba{\mathbf{a}}

\newtheorem{proposition}{Proposition}

\begin{document}

\title{Secure Wireless Communications via Cooperation}

\author{\authorblockN{Lun Dong$^\dag$, Zhu Han$^\ddag$, Athina P. Petropulu$^\dag$ and H. Vincent
Poor$^*$\\
\authorblockA{$^\dag$Electrical \& Computer Engineering Department, Drexel University\\
$^\ddag$Electrical \& Computer Engineering Department, University of Houston \\
$^*$School of Engineering and Applied Science, Princeton
University}}\thanks{This research was supported in part by the
National Science Foundation under Grants ANI-03-38807, CNS-06-25637
and
CCF-07-28208.}}

\maketitle

\begin{abstract}
The feasibility of physical-layer-based security approaches for
wireless communications in the presence of one or more eavesdroppers
is hampered by channel conditions. In this paper, cooperation is
investigated as an approach to overcome this problem and improve the
performance of secure communications. In particular, a
decode-and-forward (DF) based cooperative protocol is considered,
and the objective is to design the system for secrecy capacity
maximization or transmit power minimization. System design for the
DF-based cooperative protocol is first studied by assuming the
availability of global channel state information (CSI). For the case
of one eavesdropper, an iterative scheme is proposed to obtain the
optimal solution for the problem of transmit power minimization. For
the case of multiple eavesdroppers, the problem of secrecy capacity
maximization or transmit power minimization is in general
intractable. Suboptimal system design is proposed by adding an
additional constraint, i.e., the complete nulling of signals at all
eavesdroppers, which yields simple closed-form solutions for the
aforementioned two problems. Then, the impact of imperfect CSI of
eavesdroppers on system design is studied, in which the ergodic
secrecy capacity is of interest.

\end{abstract}

\section{Introduction}
Due to the broadcast nature of wireless channels, the issues of
privacy and security in wireless networks have taken on an
increasingly important role, especially in military and homeland
security applications. Physical (PHY) layer based security using an
information-theoretic point of view is attracting much attention in
this context. The basic idea of PHY-based security is to exploit the
physical characteristics of the wireless channel. In the real world,
signals transmitted over physical channels experience impairments
such as channel fading and additive noise. While channel fading and
thermal noise have traditionally been viewed as impediments, PHY
layer security approaches can exploit these channel characteristics
in order to enhance the security of digital communication systems.
This line of work was pioneered by Wyner, who introduced the wiretap
channel and established the possibility of creating almost perfectly
secure communication links without relying on private (secret) keys
\cite{Wyner}. Wyner showed that when the eavesdropper channel is a
degraded version of the main channel, the source and destination can
exchange perfectly secure messages at a non-zero rate, while the
eavesdropper can learn almost nothing about the messages from its
observations. The maximal rate of secrecy information from the
source to its intended destination is defined by the term
\emph{secrecy capacity}. Follow-up work by Leung-Yan-Cheong and
Hellman characterized the secrecy capacity of scalar Gaussian
wire-tap channel \cite{Leung-Yan-Cheong}. In a further paper,
Csisz\'{a}r and K\"{o}rner generalized Wyner's approach by
considering the transmission of confidential messages over broadcast
channels \cite{Csiszar}. Recently, there have been considerable
efforts devoted to generalizing these studies to the wireless
channel and multi-user scenarios (see \cite{Lai}-\cite{Gopala} and
references therein).

The feasibility of traditional PHY-based security approaches based
on single antenna systems is hampered by channel conditions: if the
channel between source and destination is worse than the channel
between source and eavesdropper, the secrecy capacity is typical
zero \cite{Wyner},\cite{Leung-Yan-Cheong}. Some recent work has been
proposed to overcome this limitation by taking advantage of multiple
antenna systems, e.g., multiple-input multiple-output (MIMO)
\cite{Hero},\cite{Negi}, single-input multiple-output (SIMO)
\cite{Parada} and multiple-input single-output (MISO)
\cite{Li:2007},\cite{Shafiee}. However, due to cost and size
limitations, multiple antennas may not be available at network
nodes. Under such scenarios, node cooperation is an effective way to
enable single-antenna nodes to enjoy the benefits of
multiple-antenna systems \cite{Laneman2}.

In this paper, we consider a situation in which each network node is
equipped with only a single omni-directional antenna and there are
one or more eavesdroppers in the network. Secure communication is
achieved via node cooperation in a decode-and-forward (DF) fashion.
We assume that source and relays are located in the same cluster,
while destination and eavesdropper(s) are at faraway locations
outside this cluster. We propose a two-stage cooperative protocol.
In Stage 1, the source node broadcasts its message locally to other
nodes within the cluster. These local transmissions typically
require a small amount of power only, and the information rate at
faraway eavesdropper(s) can be ignored. Thus, transmissions in Stage
1 can be considered to be secure. In Stage 2, relay nodes decode the
received messages. Then, the source node and relay nodes
cooperatively transmit a weighted version of the message signal to
the destination.

Our focus is on secret communications in Stage 2. We are interested
in two optimization problems: (1) design node weights to maximize
the secrecy capacity for a fixed transmit power; and (2) design node
weights to minimize the transmit power for a fixed secrecy capacity.
We assume that the global channel state information (CSI) is
available for weight design. Cooperation is here used in place of
multiple transmit antennas in MISO systems. Since there is a step
involved before transmission, during which the information is made
available to the relays, the corresponding secrecy capacity is half
of that corresponding to a MISO system. We should also point out
that existing results for system design for a centralized MISO
system can be also applied in system design for DF-based cooperative
protocols. For example, in the case of one eavesdropper, the
closed-form expression for weights that maximize the secrecy
capacity subject to a transmit power constraint has been studied in
\cite{Li:2007}, \cite{Shafiee}. Beyond existing results in
\cite{Li:2007},\cite{Shafiee}, we here propose the following new
results for the DF-based cooperative protocol: (1) For the case of
one eavesdropper, we study system design to minimize the transmit
power for a fixed secrecy capacity. We propose an iterative
algorithm to reach the optimal solution, by using the solution for
the problem of maximizing the secrecy capacity for a fixed transmit
power. (2) Prior work considered the presence of one eavesdropper
only. For the case of multiple eavesdroppers, the aforementioned
optimization problems are in general intractable. We obtain a
suboptimal (in terms of secrecy capacity or transmit power) but
simple closed-form solution, by introducing an additional
constraint, i.e., complete nulling of signals at all eavesdroppers.
(3) Prior work assumed either complete knowledge of the
eavesdroppers' channels, or only the channel statistics. In this
paper, we investigate the weight design for the more practical case
in which only imperfect estimates of eavesdroppers' channels are
available.

This paper is organized as follows. In Section \ref{DF-protocol},
the system model and the DF-based cooperative protocol is described.
In Section \ref{systemdesign}, single and multiple eavesdroppers
cases are investigated for the secrecy capacity maximization problem
and the power minimization problem. The case of imperfect CSI of
eavesdroppers is also studied. Simulations are described in Section
\ref{sec:simulation}, and conclusions are drawn in Section
\ref{sec:conclusion}.

We adopt the following notation. Bold uppercase letters denote
matrices and bold lowercase letters denote column vectors. Transpose
and conjugate transpose are represented by $(\cdot)^T$ and
$(\cdot)^\dag$ respectively; $\bI_M$ is the identity matrix of size
$M \times M$; $\mathrm{diag} \{ \ba \}$ denotes a diagonal matrix
with the elements of vector $\ba$ along its diagonal; $\mathbf{0}_{M
\times N}$ denotes an all-zero matrix of size $M \times N$;
$\mathcal{CN}(\mu,\sigma^2)$ denotes circularly symmetric, complex
Gaussian distribution with mean $\mu$ and variance $\sigma^2$;
$\bE\{\cdot\}$ denotes expectation.

\section{System Model and Cooperative Protocol} \label{DF-protocol}

\subsection{System Model}

We consider a wireless network model consisting of one source node
(node index: 0), $N-1$ ($N>1$) trusted relay nodes (node index $1,
2,\ldots, N-1$), a destination node, and $J$ ($J \geq 1$)
eavesdroppers. We assume that the source and relays are located
within the same cluster, while the destination and eavesdropper(s)
are at faraway locations from this cluster. Each node is equipped
with a single omni-directional antenna and operates in half-duplex
mode.

\begin{figure}[htb]
 \centerline{\epsfig{figure=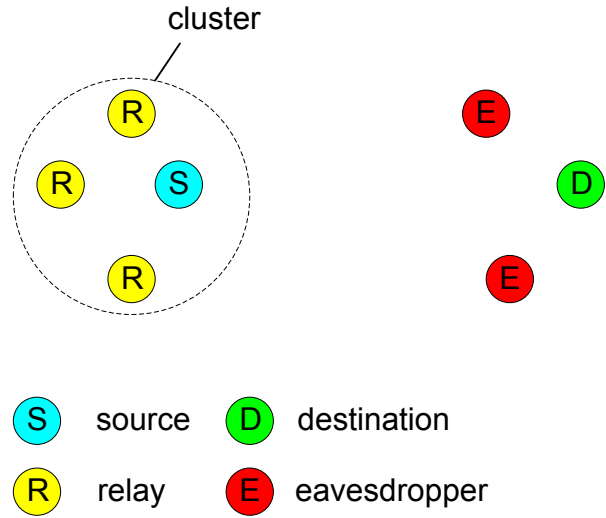,width=8cm}}
\caption{System model in the presence of eavesdroppers.}
 \label{securesystemmodel}
\end{figure}

A narrowband message signal $s_0$ is to be transmitted from the
source to the destination. The power of the message signal $s_0$ is
normalized to one, i.e, $\mathbb{E}\{|s_0|^2\} =1$. All channels are
flat fading. Let $h_i$ denote the baseband complex channel gain
between the $i$th cluster node and the destination, and $g_{i,j}$
denote the channel gain between the $i$th cluster node and the $j$th
eavesdropper. Thermal noise at all nodes is assumed to be zero-mean
white complex Gaussian, i.e., $\mathcal{CN}(0,\sigma^2)$. The
configuration is illustrated in Fig. \ref{securesystemmodel}.

We assume that the global CSI is available for system design (the
same assumption as in most of PHY-based security literature). In
practice, destination-related CSI can be obtained by periodic
pilots, and eavesdroppers-related CSI and the number of
eavesdroppers may be obtained by monitoring the behavior of
eavesdroppers. A cluster head (CH) then collects the global CSI,
executes the weight computation algorithm and sends the weights back
to cluster nodes for cooperative transmissions.

A DF-based cooperative protocol will be used. The number of relays
with successful decoding is assumed to be known a priori (rather
than being a random variable). To implement this in practice, each
relay with successful decoding can send a non-interfering
notification message to the CH.

\subsection{Cooperative Protocol} \label{protocols}

In this subsection, we describe the DF-based cooperative
transmission protocol based on our system model.

\textbf{Stage 1}: The source broadcasts its message signal $s_0$
locally to its trusted relays within the cluster. The transmit power
is chosen so that the signal $s_0$ can be decoded at the relays with
high probability. In this paper, for simplicity we assume that the
transmit power in Stage 1 is known a priori.

This stage usually requires a small amount of power only, and the
information rate at the faraway eavesdropper(s) can be ignored.
Thus, transmissions in Stage 1 can be considered to be secure.

\textbf{Stage 2}:

All the trusted relays that successfully decode the message $s_0$,
together with the source, cooperatively transmit signal $s_0$ to the
destination. For convenience, we assume that all the $N-1$ relays
successfully decode the message signal\footnote{the case in which $M
< N-1$ relays successfully decode the message is equivalent to the
case in which the total number of relays is $M$.}. Then, totally $N$
nodes ($N-1$ relays plus one source), indexed by $i=0,\ldots,N-1$,
participate in cooperative transmissions in Stage 2. Specifically,
the $i$th node transmits a weighted signal of $s_0$, i.e., $w_i
s_0$, $i=0,\ldots,N-1$, where $w_i$ is the weight of the $i$th node.

Let us define the $N \times 1$ vectors $\bw =
[w_0,\ldots,w_{N-1}]^H$, $\bh = [h_0, \ldots, h_{N-1}]^H$ and
$\bg_{j} = [g_{0,j}, \ldots, g_{N-1,j}]^H$, and the $N \times N$
matrices $\bR_{\mathrm{h}} = \bh \bh^H$ and $\bR_{\mathrm{g}}^{j} =
\bg_{j} \bg_{j}^H$.

At the destination, the received signal $y_d$ equals
\begin{eqnarray}
y_d &=& \bw^H \bh s_0 + n_d \ ,
\end{eqnarray}
where $n_d$ represents white complex Gaussian noise at the
destination. Then, the capacity at the destination is
\begin{eqnarray} \label{CapDesDF}
C_d = \frac{1}{2} \log_2 \left(1+\frac{\bw^H \bR_{\mathrm{h}}
\bw}{\sigma^2}\right)
\end{eqnarray}
where the scalar factor $1/2$ is due to the fact that two time units
are required in the two-stage cooperative protocol.

At the $j$th eavesdropper, the received signal $y_e^j$ equals
\begin{eqnarray}
y_e^j &=& \bw^H \bg_j s_0 + n_e^j \ ,
\end{eqnarray}
where $n_e^j$ represents white complex Gaussian noise at the $j$th
eavesdropper. The capacity at the $j$th eavesdropper is then
\begin{eqnarray} \label{CapEavDF}
C_e^j = \frac{1}{2} \log_2\left(1+ \frac{\bw^H \bR_{\mathrm{g}}^j
\bw}{\sigma^2}\right) \ .
\end{eqnarray}

Our objective is to design the node weights to maximize secrecy
capacity for a fixed transmit power, or minimize transmit power for
a fixed secrecy capacity. The secrecy capacity for $J$ eavesdroppers
is defined as \cite{Liang:June2008}:
\begin{eqnarray} \label{SecCap}
C_s = \max\{0,C_d-\max(C_e^1,\ldots,C_e^J)\} \ .
\end{eqnarray}

\section{System Design for Secure Wireless Communications}
\label{systemdesign}

In this section, we discuss the weight design for the DF-based
cooperative protocol to achieve secure wireless communications, for
the cases of one eavesdropper and multiple eavesdroppers,
respectively.

\subsection{One Eavesdropper} We first discuss the simple
scenario of one eavesdropper. For notational convenience, the index
of the eavesdropper is dropped. As long as $\bh \neq \bg$, we can
always find a set of weights so that the secrecy capacity is
non-zero. For example, one can completely null out the received
signal at the eavesdropper. Thus, from (\ref{CapDesDF}) and
(\ref{CapEavDF}), Eq. (\ref{SecCap}) can be written as
\begin{eqnarray}
C_s = C_d - C_e = \frac{1}{2} \log_2 \left( \frac{\sigma^2 + \bw^H
\bR_{\mathrm{h}} \bw}{\sigma^2 + \bw^H \bR_{\mathrm{g}} \bw} \right)
\ .
\end{eqnarray}

\subsubsection{Maximizing Secrecy Capacity for Fixed Transmit
Power} \label{fixpower_single}

The problem of maximizing the secrecy capacity $C_s$ for a fixed
transmit power $\bw^H \bw = P_0$ can be formulated as
\begin{eqnarray} \label{sec_opt1}
& \arg \max\limits_{\bw} \ \frac{\sigma^2 + \bw^H
\bR_{\mathrm{h}} \bw}{\sigma^2 + \bw^H \bR_{\mathrm{g}} \bw} & \\
\nonumber & \mathrm{s.t.} \ \bw^H \bw = P_0 \ . &
\end{eqnarray}
The solution of this Rayleigh quotient problem, reported in
\cite{Li:2007},\cite{Shafiee}, is the scaled eigenvector
corresponding to the largest eigenvalue  of the symmetric matrix
$\widetilde{\bR}_{\mathrm{g}}^{-1} \widetilde{\bR}_{\mathrm{h}}$,
where
\begin{eqnarray}
\widetilde{\bR}_{\mathrm{h}} \triangleq \frac{\sigma^2}{ P_0}
\bI_{N} + \bR_{\mathrm{h}}
\end{eqnarray}
and
\begin{eqnarray} \label{tildeG}
\widetilde{\bR}_{\mathrm{g}} \triangleq \frac{\sigma^2}{P_0} \bI_N +
\bR_{\mathrm{g}} \ .
\end{eqnarray}
Also, the equality power constraint in (\ref{sec_opt1}) is
equivalent to the inequality power constraint $\bw^H \bw \leq P_0$
\cite{Li:2007},\cite{Shafiee}. As we will show in the next
subsection, the solution of the problem in (\ref{sec_opt1}) can help
solve another optimization problem of minimizing transmit power
under a fixed secrecy capacity.

\subsubsection{Minimizing Transmit Power for Fixed Secrecy
Capacity} \label{MinPower_Single} The problem of minimizing the
transmit power $\bw^H \bw$ for a fixed secrecy capacity $C_s^0 > 0$
can be formulated as
\begin{eqnarray} \label{sec_opt3}
& \arg \min\limits_{\bw} \bw^H \bw &  \\
& \mathrm{s.t.} \ \frac{\sigma^2 + \bw^H \bR_{\mathrm{h}}
\bw}{\sigma^2 + \bw^H \bR_{\mathrm{g}} \bw} = 4^{C_s^0} & \ .
\nonumber
\end{eqnarray}
However, the conventional method of Lagrange multipliers does not
work for (\ref{sec_opt3}), as it yields a zero solution of $\bw$. To
solve (\ref{sec_opt3}), we first propose the following.

\begin{proposition}\label{proposition1}
The solutions of the following two optimization problems are
identical:

(i) Find the weights that maximize $C_s$ for a fixed transmit power
$P_0$.

(ii) Find the weights that minimize the transmit power for a fixed
$C_s^{\max}$, where $C_s^{\max}$ is the maximal $C_s$ of problem
(i).

\begin{proof}
We prove Proposition \ref{proposition1} by contradiction. We assume
that $\bw^{(1)}$ is the optimal solution that yields $C_s^{(1)} =
C_s^{\max}$ for fixed transmit power $P_0$, while a different weight
vector $\bw^{(2)} \neq \bw^{\mathrm{(1)}}$ minimizes the transmit
power for fixed $C_s^{(2)}= C_s^{\max}$. Thus, the transmit power
$(\bw^{(2)})^H \bw^{(2)}$ must be smaller than $(\bw^{(1)})^H
\bw^{(1)} = P_0$. We can always find a scalar $\rho
>1$ such that the weight vector $\rho \cdot
\bw^{(2)}$ also achieves $\rho^2 (\bw^{(2)})^H \bw^{(2)} = P_0$.

Now, we prove that $C_s$ based on the weight vector $\rho \cdot
\bw^{(2)}$ is greater than $C_s^{\max}$. Let us define the function
\begin{eqnarray}
F(z) = \frac{\sigma^2 + z^2 (\bw^{(2)})^H \bR_{\mathrm{h}}
\bw^{(2)}}{\sigma^2 + z^2(\bw^{(2)})^H \bR_{\mathrm{g}} \bw^{(2)}} \
.
\end{eqnarray}
We can equivalently prove $F(\rho) > F(1)$ for $\rho>1$. Taking the
derivative of $F(z)$ with respect to $z$, we obtain
\begin{eqnarray}
\frac{d F(z)}{d z} \propto (\bw^{(2)})^H \bR_{\mathrm{h}} \bw^{(2)}
- (\bw^{(2)})^H \bR_{\mathrm{g}} \bw^{(2)}.
\end{eqnarray}
As $C_s>0$, $\frac{d F(z)}{d z} > 0$. Thus, $F(z)$ is a
monotonically increasing function of $z$ and it follows that
$F(\rho)
> F(1)$ for $\rho>1$.
Hence, we have proved that $C_s$ based on the weight vector $\rho
\cdot \bw^{(2)}$ is greater than $C_s^{\max}$. In other words,
$C_s^{\max}$ is not the maximal value of $C_s$ for transmit power
$P_0$, which contradicts our assumption. Therefore, $\bw^{(1)}$ must
be equal to $\bw^{(2)}$, and thus Proposition \ref{proposition1} is
proved.
\end{proof}
\end{proposition}

Based on Proposition \ref{proposition1}, we now propose the
following iterative algorithm for finding the optimal solution of
(\ref{sec_opt3}).

\begin{itemize}
\item \textbf{Initialization:} \\
S0) Set an initial value for the weights $\rho^{(0)} \bw^{(0)}$,
where $\rho^{(0)}$ is a scalar such that $C_s$ for $\bw^{(0)}$
equals $C_s^{0}$. Note that $\bw^{(0)}$ can be arbitrarily chosen
but its corresponding secrecy capacity must be greater than zero.
Then, compute the transmit power $P^{(0)} = (\rho^{(0)})^2
(\bw^{(0)})^H \bw^{(0)}$.

\item \textbf{Iteration:} \\
S1) In the $k$th iteration, compute the weight vector $\bw^{(k)}$
that maximizes the secrecy capacity for fixed transmit power
$P^{(k-1)}$, based on the method discussed in Section \ref{fixpower_single}. \\
S2) Compute the scalar $\rho^{(k)}$, such that $C_s$ under
$\rho^{(k)} \bw^{(k)}$ equals $C_s^{0}$. Calculate the updated
transmit power $P^{(k)}=(\rho^{(k)})^2 (\bw^{(k)})^H
\bw^{(k)}$. \\
S3) Iterate until $P^{(k-1)} - P^{(k)}$ is smaller than a
pre-defined threshold.
\end{itemize}

The objective function of (\ref{sec_opt3}) is convex and the updated
power with each iteration is nonincreasing. Thus, the above
algorithm eventually converges to the global minimum. In our
simulations, the iteration always converged very rapidly.

\subsection{Multiple Eavesdroppers} \label{multinodes}
In this subsection we discuss the scenario of $J>1$ eavesdroppers.
From (\ref{SecCap}), the secrecy capacity for multiple eavesdroppers
is related to the capacity at all eavesdroppers. Determining the
weights that maximize secrecy capacity for fixed power, or minimize
power for fixed secrecy capacity is in general intractable. In the
following, we consider an additional constraint, i.e., completely
nulling out signals at all eavesdroppers. The resulting secrecy
capacity (transmit power) represents a lower (upper) bound of the
optimal one.

\subsubsection{Minimizing Transmit Power for Fixed Secrecy Capacity}
\label{MinPower_multi} Let us define the $N \times J$ matrix
$\bG=[\bg_1,\ldots,\bg_J]$. To null the signals at all
eavesdroppers, we need
\begin{eqnarray} \label{nulls}
 \bw^H \bG = \mathbf{0}_{1 \times J} \ .
\end{eqnarray}

To satisfy the fixed secrecy capacity $C_s^0$, we also need
\begin{eqnarray} \label{desired1}
C_s^0 = C_d = \frac{1}{2} \log_2 \left(1+\frac{\bw^H
\bR_{\mathrm{h}} \bw}{\sigma^2}\right) \ .
\end{eqnarray}
Eq. (\ref{desired1}) can also be written as
\begin{eqnarray} \label{desired2}
\bw^H \bh = \sqrt{(4^{C_s^0}-1)\sigma^2} \cdot e^{j \theta}
\end{eqnarray}
where $\theta$ is an arbitrary angle within $[0,2\pi)$.

Defining the $(J+1) \times N$ matrix $\widetilde{\bG} = [\bh,\bG]^H$
and the $(J+1) \times 1$ vector $\be = [1, \mathbf{0}_{1 \times
J}]^T$, we can rewrite the constraints in (\ref{nulls}) and
(\ref{desired2}) as
\begin{eqnarray} \label{constraints}
 \widetilde{\bG} \bw = (\sqrt{(4^{C_s^0}-1)\sigma^2} \cdot e^{j \theta}) \be \ .
\end{eqnarray}
To guarantee a non-zero solution for $\bw$, we need $N \geq J+1$,
which usually can be easily satisfied.

The optimal solution $\bw^{\mathrm{opt}}$ that minimizes the
transmit power corresponds to the least-squares solution of
(\ref{constraints}) produced by the pseudo-inverse of
$\widetilde{\bG}$ \cite{Boyd},\cite{Hanbook}, i.e.,
\begin{eqnarray} \label{weights}
\bw^{\mathrm{opt}} = (\sqrt{(4^{C_s^0}-1)\sigma^2} e^{j \theta})
\widetilde{\bG}^H (\widetilde{\bG} \widetilde{\bG}^H)^{-1} \be \ .
\end{eqnarray}

From (\ref{weights}), the transmit power $(\bw^{\mathrm{opt}})^H
\bw^{\mathrm{opt}}$ is independent of the selection of $\theta$. For
convenience we can take $\theta=0$.

\subsubsection{Maximizing Secrecy Capacity for Fixed Transmit Power}
\label{fixPower_multi}

The optimization problem can be formulated as
\begin{eqnarray} \label{sec_opt4}
& \arg \max\limits_{\bw} \bw^H \bR_{\mathrm{h}}
\bw & \\
& \mathrm{s.t.} \ \nonumber  \bw^H \bw = P_0 \ \mbox{and} \ \bw^H
\bG = \mathbf{0}_{1 \times J}  \ . &
\end{eqnarray}

The conventional method of Lagrange multipliers does not yield an
insightful closed-form solution of (\ref{sec_opt4}). To solve
(\ref{sec_opt4}), we propose the following.

\begin{proposition}\label{proposition2}
The solutions of the following two optimization problems are
identical:

(i) Find the weights that maximize $C_s$ for fixed transmit power
$P_0$, and also meets the constraint that signals at all
eavesdroppers are completely nulled. Let us denote the maximal $C_s$
by $C_s^{\max}$.

(ii) Find the weights that minimize the transmit power for a fixed
$C_s^{\max}$ and also meets the constraint that signals at all
eavesdroppers are completely nulled.

\begin{proof}
We follow arguments similar to those used in the proof of
Proposition \ref{proposition1}. We assume that weight vector
$\bw^{(1)}$ achieves $C_s^{\max}$ for  the fixed transmit power
$P_0$, while a different weight vector $\bw^{(2)}\neq \bw^{(1)}$
achieves minimal transmit power for fixed $C_s^{\max}$. Thus, it
holds that $(\bw^{(2)})^H\bw^{(2)} < P_0$. We can always find a
scalar $\rho > 1$ such that under the weights $\rho \cdot \bw^{(2)}$
the transmit power is $\rho^2 (\bw_\ell^{(2)})^H\bw_\ell^{(2)} =
P_0$. However, the weight vector $\rho \cdot \bw^{(2)}$ achieves a
secrecy capacity greater than $C_s^{\max}$. In other words,
$\bw^{(1)}$ does not achieve the maximum of $C_s$ for fixed power
$P_0$, which contradicts our assumption. Therefore, $\bw^{(1)}$ must
be equal to $\bw^{(2)}$.
\end{proof}
\end{proposition}

From Proposition \ref{proposition2}, the optimization problem of
(\ref{sec_opt4}) is equivalent to finding the weights that minimize
the transmit power for fixed $C_s^{\mathrm{max}}$. From
(\ref{weights}), the transmit power is proportional to
$4^{C_s^0}-1$. Thus, the solution of (\ref{sec_opt4}) is
\begin{eqnarray} \label{sol_multi_fixedcap}
\bw^{\mathrm{opt}} = \beta \widetilde{\bG}^H (\widetilde{\bG}
\widetilde{\bG}^H)^{-1} \be
\end{eqnarray}
where $\beta$ is a scalar and equals
\begin{eqnarray}
\beta = \sqrt{\frac{P_0}{\be^H (\widetilde{\bG}
\widetilde{\bG}^H)^{-1} \be}} \ .
\end{eqnarray}

Substituting (\ref{sol_multi_fixedcap}) into the objective function
of (\ref{sec_opt4}), one can see that the secrecy capacity is a
monotonically increasing function of the power budget $P_0$. Thus,
the equality power constraint in (\ref{sec_opt4}) is equivalent to
the inequality power constraint $\bw^H \bw \leq P_0$.

\subsection{Impact on Imperfect CSI of Eavesdroppers}
\label{imperfectCSI}

The channels between cluster nodes and the destination can be
estimated accurately, since they are trusted nodes. However, in
practice there will be some certain estimation errors for the
channels between cluster nodes and the eavesdroppers. In this
subsection, we discuss weight design for such cases.

We model the perfect channels of the $j$th eavesdropper as $\bg_j =
\widehat{\bg}_j + \Delta_j$, where $\widehat{\bg}_j$ is the
imperfect channel estimate available for weight computation, and
$\Delta_j$ corresponds to the channel error. We further assume that
the entries of $\Delta_j$ are zero-mean random variables, and
$\bR_{\Delta} \triangleq \bE \{\Delta_j \Delta_j^H \}$ is known a
priori and is independent of $j$. Thus, we obtain
\begin{eqnarray} \label{newG}
\bR_{\mathrm{g}}^j \triangleq \bE \{ \bg_j \bg_j^H \} =
\widehat{\bR}_{\mathrm{g}}^j + \bR_{\Delta}
\end{eqnarray}
where $\widehat{\bR}_{\mathrm{g}}^j = \widehat{\bg}_j
\widehat{\bg}_j^H$.

Note that we still assume the availability of perfect CSI of the
destination.

\subsubsection{One Eavesdropper} For one eavesdropper, the ergodic
secrecy capacity is given by
\begin{eqnarray}
\overline{C}_s &=& \frac{1}{2} \log_2 \left(1+\frac{\bw^H
\bR_{\mathrm{h}} \bw}{\sigma^2}\right) \nonumber\\
&& - \bE\left\{ \frac{1}{2} \log_2\left(1+ \frac{\bw^H \bg \bg^H
\bw}{\sigma^2}\right)\right\} \ .
\end{eqnarray}
The optimization problem of maximizing ergodic secrecy capacity
under a fixed power is in general difficult. To simplify the
problem, we use Jensen's inequality to obtain
\begin{eqnarray} \label{lowbound}
\overline{C}_s & \geq & \frac{1}{2} \log_2 \left(1+\frac{\bw^H
\bR_{\mathrm{h}} \bw}{\sigma^2}\right) \nonumber \\
&& - \frac{1}{2} \log_2\left(1+ \frac{ \bw^H \bR_{\mathrm{g}}
\bw}{\sigma^2}\right)
\end{eqnarray}
in which the eavesdropper index is omitted for notational
convenience. We now consider the problem of maximizing the lower
bound on ergodic secrecy capacity in (\ref{lowbound}) under a fixed
power $\bw^H \bw = P_0$. It is easy to see that this optimization
problem is the same as (\ref{sec_opt1}), while the matrix
$\bR_{\mathrm{g}}$ is now given by (\ref{newG}). Also, the problem
of minimizing the transmit power under a fixed lower bound on
ergodic secrecy capacity can be solved by the iterative algorithm in
section \ref{MinPower_Single}.

\subsubsection{Multiple Eavesdroppers} \label{imperfectCSIMultiEav}
For $J$ eavesdroppers ($J>1$), the lower bound on ergodic secrecy
capacity is given by
\begin{eqnarray}
\overline{C}_s &\geq& \frac{1}{2} \log_2 \left(1+\frac{\bw^H
\bR_{\mathrm{h}} \bw}{\sigma^2}\right) \nonumber\\
&& - \max\limits_{j} \left\{ \frac{1}{2} \log_2 \left( 1+ \frac{
\bw^H \bR_{\mathrm{g}}^j \bw}{\sigma^2} \right) \right\}
\end{eqnarray}
where $\bR_{\mathrm{g}}^j$ is given by (\ref{newG}).

To form nulls at all eavesdroppers, we need $\bw^H
\bR_{\mathrm{g}}^j \bw = 0$ for $j=1,\ldots,J$. A non-zero solution
exists only if $\bR_{\Delta}$ is semi-positive definite. In case for
which $\bR_{\Delta}$ is strictly positive definite, nulls cannot be
formed at eavesdroppers, and $\bw^H \bR_{\mathrm{g}}^j \bw$ is
always greater than zero. To cover all cases, here we still consider
the constraint $\bw^H \widehat{\bR}_{\mathrm{g}}^j \bw = 0$ or
equivalently $\bw^H \widehat{\bg}_j = 0$. The optimization problem
of maximizing the lower bound on the ergodic secrecy capacity in
(\ref{lowbound}) under a fixed power can be formulated as
\begin{eqnarray} \label{sec_opt5}
& \arg \max\limits_{\bw} \frac{\sigma^2+\bw^H \bR_{\mathrm{h}} \bw}{\sigma^2+\bw^H \bR_{\Delta} \bw} & \\
& \mathrm{s.t.} \ \nonumber  \widehat{\bG} \bw = \mathbf{0}_{J
\times 1} \ \mbox{and} \ \bw^H \bw = P_0
 &
\end{eqnarray}
where $\widehat{\bG} \triangleq
[\widehat{\bg}_1,\ldots,\widehat{\bg}_{J}]^H$. Let us define the
matrix $\bT$ containing all of the right singular vectors
corresponding to zero singular values of $\widehat{\bG}$. To satisfy
the first constraint in (\ref{sec_opt5}), $\bw$ shall be a linear
combination of basis in the null space of $\widehat{\bG}$, i.e.,
$\bw = \bT \bv$, where $\bv$ is a column vector. Then, the
optimization problem in (\ref{sec_opt5}) is equivalent to
\begin{eqnarray} \label{sec_opt6}
& \arg \max\limits_{\bv} \frac{\sigma^2+\bv^H \bT^{H} \bR_{\mathrm{h}} \bT \bv}{\sigma^2+\bv^H \bT^{H} \bR_{\Delta} \bT \bv} & \\
& \mathrm{s.t.} \nonumber \ \bv^H \bv = P_0
 &
\end{eqnarray}
which is a Rayleigh quotient problem similar to (\ref{sec_opt1}).
The final solution of (\ref{sec_opt5}) is then $\bw = \sqrt{P_0} \bT
\bq_{\mathrm{unit}}$ where $\bq_{\mathrm{unit}}$ is the unit-norm
eigenvector of the matrix $\bT^{H} [\bR_{\Delta} +
(\sigma^2/P_0)\bI]^{-1} [\bR_{\mathrm{h}} + (\sigma^2/P_0)\bI] \bT$
corresponding to its largest eigenvalue.

Due to the similarity between (\ref{sec_opt6}) and (\ref{sec_opt1}),
and the duality as shown in Proposition \ref{proposition1}, the
problem of minimizing the transmit power under a fixed lower bound
on secrecy capacity can be solved by the iterative algorithm in
section \ref{MinPower_Single}.

\subsection{Discussion}
In the above analysis, for convenience we have assumed that the
transit power in Stage 1 is much smaller than the transmit power in
Stage 2, and thus the information rates in Stage 1 at the faraway
destination and eavesdropper(s) are ignored. In this subsection, we
discuss the effects on weight design when the information rates in
Stage 1 are also taken into account.

When both stages are taken into account, the destination or an
eavesdropper combines the two received signal in both stages using
maximal ratio combining (MRC) in order to maximize the
signal-to-noise ratio (SNR). Suppose that transmit power in Stage 1
is $\tilde{P}_0$. The capacity at the destination is given by
\begin{eqnarray}
C_d &=& \frac{1}{2} \log_2 \left(\alpha +\frac{\bw^\dag
\bR_{\mathrm{a}} \bw}{\sigma^2}\right)
\end{eqnarray}
where $\alpha \triangleq 1 + \tilde{P}_0 |h_0|^2/\sigma^2$. Note
that $\tilde{P}_0 |h_0|^2/\sigma^2$ is the received SNR in Stage 1
at the destination. Similarly, the capacity at the $j$th
eavesdropper is
\begin{eqnarray}
C_e^j &=& \frac{1}{2} \log_2 \left(\mu+\frac{\bw^\dag
\bR_{\mathrm{b}}^j \bw}{\sigma^2}\right)
\end{eqnarray}
where $\mu \triangleq 1 + \tilde{P}_0  |g_{0,j}|^2/\sigma^2$.  Note
that $\tilde{P}_0 |g_{0,j}|^2/\sigma^2$ is the received SNR in Stage
1 at the $j$th eavesdropper. Here, $\alpha$ and $\mu$ are considered
to be constants, as $\tilde{P}_0$ is assumed to be a priori.

Therefore, the only change on the capacity of the destination or
eavesdropper is to replace the constant one in (\ref{CapDesDF}) or
(\ref{CapEavDF}) by $\alpha$ or $\mu$. It is easy to show that most
of the proposed analysis (when ignoring Stage 1) can still be
applied here, subject to minor changes only. The only exception is
the power minimization problem for the case of one eavesdropper (see
Section \ref{MinPower_Single}). For this case, in order to guarantee
the validation of Proposition \ref{proposition1}, the fixed secrecy
capacity $C_s^0$ should be chosen to satisfy $\mu \bw^H
\bR_{\mathrm{h}} \bw
> \alpha \bw^H \bR_{\mathrm{g}} \bw$ for every possible $\bw$.

\begin{figure}[tb]
 \centerline{\epsfig{figure= 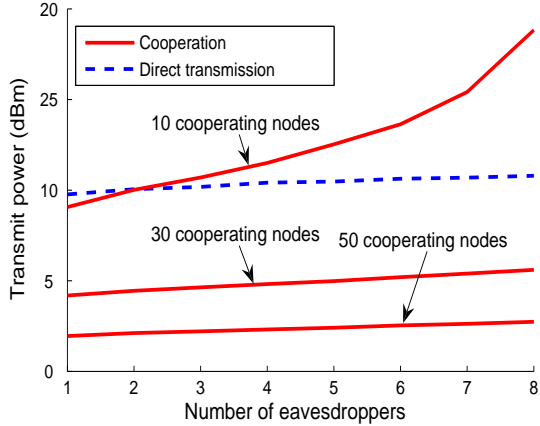,width=8.0cm,height=6.0cm}}
    \caption{Transmit power vs. number of eavesdroppers. Secrecy capacity is fixed at $C_s^0 = 3$
b/s/Hz.} \label{transmitpower1}
\end{figure}

\section{Simulations}\label{sec:simulation}
In this section, we investigate the performance of weight design
algorithms via simulations. In these simulations, the carrier
frequency is 900 MHz and the signal wavelength is $\lambda = 0.33$
m. The noise power $\sigma^2$ is $-60$ dBm. The cluster is a disk
with radius $R=5 \lambda$. The cluster nodes are uniformly located
in the disk. For convenience, a simple line-of-sight channel model
is used: $h_i = d_{i}^{-\frac{\alpha}{2}} e^{j \phi_i}$ where
$d_{i}$ is the distance between the $i$th node and the destination,
$\alpha = 4$ is the path loss exponent and $\phi_i$ denotes the
phase offset. $g_{ij}$ is defined in a similar way. All channel
estimates are assumed to be perfect.

\begin{figure}[htb]
 \centerline{\epsfig{figure=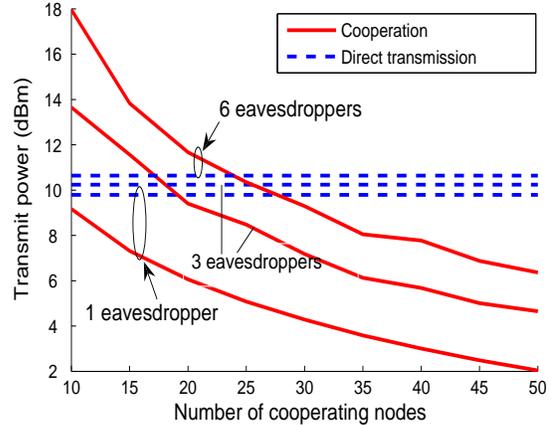,width=8.0cm,height=6.0cm}}
 \caption{ Transmit power vs. number of cooperating nodes. Secrecy capacity is fixed at $C_s^0 = 3$
b/s/Hz.}
 \label{transmitpower2}
\end{figure}

We try to compare the performance of DF-based cooperation with
direct transmission (without cooperation). Based on the
line-of-sight channel model, when the distance between any
eavesdropper and the source is smaller than the distance between the
destination and the source, the secrecy capacity of direct
transmission without cooperation is always zero no matter how large
the transmit power is. Thus, under such scenarios, cooperation
always outperforms direct transmission. In the following
simulations, we will focus on the case in which the distances
between eavesdroppers and the source are greater than the distance
between the destination and the source. The distances between the
source and destination is $20 R$. The distances between the source
and eavesdroppers are uniformly distributed within $[40R, 100R]$,
and the azimuthal directions of eavesdroppers are uniformly
distributed within $[0,2\pi)$. We perform a Monte-Carlo experiment
consisting of 1000 independent trials to obtain the average results.
Locations of cluster nodes and eavesdroppers in one trial are chosen
independently from those in other trials.

\subsection{Fixed Secrecy Capacity}
We first fix the secrecy capacity at $C_s^0=3$ b/s/Hz and
investigate the performance of transmit power. Fig.
\ref{transmitpower1} shows the transmit power versus number of
eavesdropper $J$. The number of cooperating nodes $N$ is 10, 30 or
50. For a single eavesdropper, the transmit power with cooperation
is obtained based on the iterative algorithm in Section
\ref{MinPower_Single}. For multiple eavesdroppers, the transmit
power with cooperation is computed from (\ref{weights}). As
observed, As observed, for both cooperation and direct transmission,
more transmit power would be needed as the number of eavesdroppers
increases. When the number of cooperating nodes is small,
cooperation may not outperform direct transmission (see the curve
for $N=10$ in Fig. \ref{transmitpower1}), as its transmission time
is longer. When the number of cooperating nodes is large,
cooperation requires much less transmit power than direct
transmission (see the curves for $N=30, 50$ in Fig.
\ref{transmitpower1}). Fig. \ref{transmitpower1} shows the transmit
power versus number of cooperating nodes $N$. The number of
eavesdroppers $J$ is one, three or six. As expected, the transmit
power for cooperation decreases as the number of cooperating nodes
$N$ increases, while the transmit power of direct transmission is
independent of $N$.

\begin{figure}[htb]
 \centerline{\epsfig{figure= 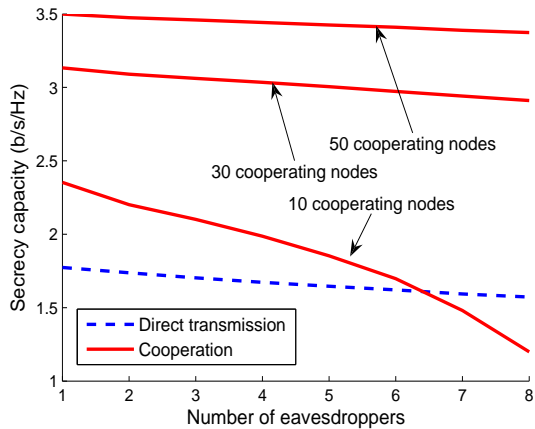,width=8.0cm,height=6.0cm}}
    \caption{Secrecy capacity vs. number of eavesdroppers. Transmit power is fixed at $P_0 = 5$
dBm.}  \label{AvgSecrecyCapacity1}
\end{figure}

\subsection{Fixed Transmit Power}
\begin{figure}[htb]
 \centerline{\epsfig{figure=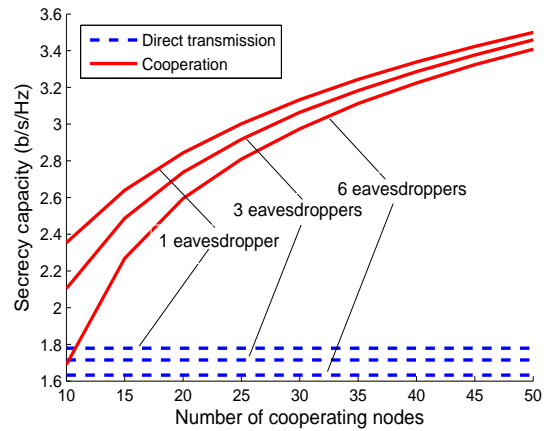,width=8.0cm,height=6.0cm}}
 \caption{Secrecy capacity vs. number of cooperating nodes. Transmit power is fixed at $P_0 = 5$
dBm. }
 \label{AvgSecrecyCapacity2}
\end{figure}

In this subsection, we investigate the performance of secrecy
capacity by fixing the transmit power at $P_0 = 5$ dBm. Fig.
\ref{AvgSecrecyCapacity1} shows the secrecy capacity versus number
of eavesdroppers. For a single eavesdropper, the secrecy capacity
with cooperation is obtained based on the result in Section
\ref{fixpower_single}. For multiple eavesdroppers, the secrecy
capacity with cooperation is computed based on the nulling weights
of (\ref{sol_multi_fixedcap}). As expected, the secrecy capacity
decreases as the number of eavesdroppers increases. A larger number
of cooperating nodes yields higher secrecy capacity. Fig.
\ref{AvgSecrecyCapacity2} shows the secrecy capacity versus number
of cooperating nodes $N$. The secrecy capacity for cooperation
increases as $N$ increases, while the secrecy capacity of direct
transmission is independent of $N$.

\section{Conclusions}\label{sec:conclusion}
In this paper, we have considered a DF-based cooperative protocol to
improve the performance of secure wireless communications in the
presence of one or more eavesdroppers. For the case of one
eavesdropper, we have considered the design problem of transmit
power minimization and have proposed an iterative algorithm to reach
the solution, by the help of existing results for another problem of
secrecy capacity maximization. For the case of multiple
eavesdroppers, we have derived suboptimal and closed-form solutions
for the problems of transmit power minimization and secrecy capacity
maximization by adding an additional constraint, i.e., the complete
nulling of signals at all eavesdroppers. We have also investigated
the impact of imperfect CSI of eavesdroppers on system design.

\end{document}